# Diamond-Like Carbon Coatings on Plasma Nitrided M2 Steel: effect of deposition parameters on adhesion properties.

A. Moreno-Bárcenas[1, 2], J.M Alvarado-Orozco[1, 3], J.M. González Carmona[1, 3], G.C. Mondragón-Rodríguez[1, 3], J. González-Hernández[1], A. García-García[2*]

[1]Center of Engineering and Industrial Development, CIDESI-Querétaro, Surface Engineering Department. Av. Pie de la Cuesta 702, Desarrollo San Pablo, Querétaro, México.

[2]Laboratorio de Síntesis y Modificación de Nanoestructuras y Materiales Bidimensionales. Centro de Investigación en Materiales Avanzados S.C., Parque PIIT, Km 10 Autopista Monterrey-Aeropuerto, Apodaca N.L. C.P. 66628, México.

[3]CONACyT, Consorcio de Manufactura Aditiva, CONMAD, Av. Pie de la Cuesta 702, Desarrollo San Pablo, Querétaro, México.

**Abstract**

Diamond-like carbon (DLC) coatings have excellent mechanical and tribological properties, as well as good wear and corrosion resistance. They are well established in medical, metalworking and automotive applications. However, further improvements are needed and substrate pre-treatment plays an important role in supporting and enforcing the adhesion and service performance of the DLC coatings. In this work, the synergetic effect of low-pressure arc plasma-assisted-nitriding (PAN) treatment of M2 steel and plasma-enhanced chemical vapor deposition (PECVD) on the DLC coatings adhesion was analyzed. Adhesion strength of a duplex DLC + nitriding coating system was compared to the performance of DLC-coatings applied on non-nitrided M2 steel substrates. The nitrided layer was analyzed by optical microscopy, X-ray

*Corresponding author: alejandra.garcia@cimav.edu.mx

The authors declare that there is not conflict of interest regarding the publication of this paper.

diffraction, Vickers microhardness and atomic force microscopy. DLC coatings were analyzed using Raman spectroscopy revealing that DLC a-C:H type was obtained. The adhesion properties were analyzed by scratch testing supported by optical microscopy and Scanning Electron Microscope. Results showed an improvement of the DLC adhesion on the plasma nitrided surfaces.

**1. Introduction**

Diamond-like carbon (DLC) coatings are well known for their extraordinary mechanical performance. Due to their high hardness, good wear resistance and low friction coefficient [1–3] DLC-coatings have been continuously applied in medical, metalworking and automotive sectors in the last years [4,5]. It is well known that the performance of the DLC films is dependent on $sp^3$ hybridizations and hydrogen content.

Ferrari and Robertson *et al*. [6] classified hydrogenated amorphous carbons into four classes: i) Polymer-like (H:C) with 60 % $sp^3$ hybridization and the highest hydrogen content (40–50 %); ii) Diamond-like (a-C:H) with intermediate hydrogen (20–40 %) and ~40 % $sp^3$ contents but still maintaining good mechanical properties; iii) Hydrogenated tetrahedral amorphous carbon (ta-C:H) with the highest $sp^3$ content (~70 %) and around 25 % of hydrogen and tetrahedral amorphous carbon (ta-C) with a low hydrogen content (<10 %) and iv) Graphite-like (g-C:H) with a low hydrogen content (less than 20 %) and high $sp^2$ proportion.

Several deposition methods have been explored in order to develop a-C:H films with higher proportions of $sp^3$ carbon atoms including sputtering and cathodic arc processes. Films grown by plasma-enhanced chemical vapor deposition (PECVD), generally present 70 to 100 % of $sp^2$ and $sp^3$ hybridizations with hydrogen contents lower than 30 at.% [7,8].

*Corresponding author: alejandra.garcia@cimav.edu.mx

On the other hand, the DLC-coatings have adhesion problems when they are deposited on metallic substrates because carbon diffuses into the metals delaying the DLC nucleation. Moreover, the thermal expansion coefficients of coating and steels are not compatible, which causes poor adhesion [9].

Besides the coating properties themselves, the substrate also has an important influence on the surface coating attributes. Well-directed surface modifications of the substrate can lead to better adhesion and longer durability of coated parts and components. Efforts have been done to develop coatings architectures that improve the adhesion of DLC-coatings on tool steels [10]. For instance, the duplex system (nitrided layer + DLC coating) can be then considered as a good option not only to enhance adhesion, but also the fretting wear resistance corrosion and fatigue resistance. Zeng et al. [11] improved the mechanical and tribological properties of a M2 steel by combining a Cr interface between the substrate and DLC coating. This combination results in improved an adhesion, load-bearing capacities and wear resistance of the DLC coatings. Other interfaces such as AlTiN [12], Ti-Al[13], and Zr [14] have been used to further improve the mechanical properties of M2 steel. Plasma assisted nitriding of tool steels produces a hardness gradient through the substrate as a result of nitrogen diffusion. The nitriding layer acts as a mechanical buffer, increases its surface hardness, improves its wear and corrosion resistance, and may increase the adhesion strength of the DLC coating [12].

*Corresponding author: alejandra.garcia@cimav.edu.mx

Currently three adhesion mechanisms between the coating-substrate systems are well accepted including mechanical interlocking, physical and chemical bonding. Mostly but not always, one of these mechanisms acts as the factor that controls adhesion. Although these adhesion mechanisms are well accepted by the scientific community, there is still a lack of understanding on the effect that nitriding has over the DLC-coating adhesion on tool steels and its controlling mechanisms. The aim of this work is to investigate the effect that a low-pressure arc plasma-assisted-nitriding (PAN) process has on the adhesion of DLC-coatings deposited at different bias voltages. The results are compared and discussed with DLC-coatings deposited on non-nitrided and nitrided M2 tool steel.

**2. Experimental details**

**2.1 Plasma assisted nitriding process**

The AISI M2 tool steel commonly used in the manufacture of a large number of cutting tools [15–17], was chosen in this investigation. Annealed steel pieces of 2.54 cm in diameter and 5 mm thickness were used as substrates. The M2 steel samples were mirror polished by applying different grades of SiC paper, diamond spray and colloidal silica. The non-nitrided control samples were only mirror polished and then ultrasonically cleaned in two steps, first with distilled water and after that in isopropanol followed by drying. A PAN process was carried out in a Domino Mini coater unit from Oerlikon. The M2 steel substrates were fixed in a planetary system with two-fold rotation and introduced inside the coating chamber. The process temperature was maintained at ~330 °C with a 9 kW pre-heater systems for 1 h. The samples were cleaned using $Ar^+$ ions during 15 min. Then, a nitriding step was maintained for 1.5 hrs and a gas mixture of 278 sccm Ar + 50 sccm $N_2$. After nitriding the substrates were cooled down to near room temperature before opening the PVD chamber.

**2.2 DLC deposition process**

The DLC-coatings were deposited on nitrided and non-nitrided M2 steels using a commercial $H_2$ CVD-PECVD reactor from Intercovamex. Prior the DLC deposits, the substrate´s surface was activated with an Ar plasma for 1 min at 100 W. Process power (W), which is related to the bias voltage ($V_{bias}$) was selected at 50, 70, 100, 150 and 200 W. The effects of the power and bias voltages on the coating adhesion and their chemical qualities were evaluated. The total gas mixture of 100 sccm (8 sccm Ar + 2 sccm $CH_4$ + 90 sccm Ar:$CH_4$) was kept constant. Deposition times were fixed at 40 min at 200 °C, table I summarizes the experimental conditions.

Table I Experimental conditions to growth DLC on nitrided and non-nitrided substrates.

| W | $V_{bias}$ | T(°C) | t(min) | N-Nitrided | NN-Non nitrided |
|---|---|---|---|---|---|
| 50 | 380 | 150 | 40 | N-50 | NN-50 |
| 70 | 450 | 150 | 40 | N-70 | NN-70 |
| 100 | 540 | 150 | 40 | N-100 | NN-100 |
| 150 | 656 | 150 | 40 | N-150 | NN-150 |
| 200 | 743 | 150 | 40 | N-200 | NN-200 |

**2.3 Characterization methods**

**2.3.1 Optical microscopy**

The microstructures, scratch surfaces, and cross-sections were observed using optical microscopy Olympus model GX1, with stream essential image analyzer to identify microstructures.

*Corresponding author: alejandra.garcia@cimav.edu.mx

**2.3.2 SEM and EDX**

Initial chemistry analyses and scratch surface morphology were obtained by a combination of Scanning Electron Microscopy (SEM) and Energy Dispersive X-ray Spectroscopy (EDX), using a FEI Company model Nova NanoSem 200 Scanning Electron Microscope, with Detector EDX Oxford model INCA x-sight.

**2.3.3 X-ray diffraction**

X-ray diffraction (XRD) was used to investigate the crystalline phases of nitrided and non-nitrided tool steel. Powder XRD patterns were recorded by a Panalytical model Empyrean diffractometer using Cu Ka radiation (45 kV, 40 mA), over a 2θ range from 35 to 90°.

**2.3.4 Microhardness**

The nitriding depth was evaluated using a Matsuzawa microdurometer model MMT-X7, with CLEMEX image analyzer for the measurement of the indentations.

**2.3.5 Raman spectra**

Coatings were recorded using a Horiba in via micro-Raman spectrometer model LabRam HR evolution with an electronically cooled (-60 C) CCD camera and a Sincerities microscope. The excitation source used was an Nd-YAG laser with emission at 532 nm, integration time was 3 seconds. A diffraction grating at 600 lines/mm and objective of 100X magnification were used.

**2.3.6 Atomic force microscope (AFM)**

Surface roughness measurements were made using an atomic force microscope (AFM) was used, Asylum, model MFP3D-SA, using the technique of noncontact mode or AC, with rectangular cantilever model AC240TS-R3 at 2N/m and a scanning speed of 70 kHz.

*Corresponding author: alejandra.garcia@cimav.edu.mx

The authors declare that there is not conflict of interest regarding the publication of this paper.

**2.3.7 Scratch tests**

The scratch tests were carried out on a BRUKER UMT-3, with a crescent load and a ball diameter of 200 μm. Three scratch tests were carried out on each coating with adjacent tracks separated by minimum 100 μm. Scratch lengths were arbitrarily chosen, while the normal load was steadily raised to ensure coating failure. Moreover, optical micrographs were used to determine the critical loads and evaluate the scratch mechanical behavior.

**3 Results and discussion**

**3.1 Optical and chemical analysis of surfaces.**

Chemical analysis by energy-dispersive X-ray spectroscopy (EDX) gave the following major alloying elements of the M2 steel (wt. %): C 0.85, W 6.09, Mo 4.76, Cr 4.05, V 1.68, Co 0.44. Fig. 1a shows the microstructure of the substrate corresponding to AISI M2 steel in the annealed state, constituted by metal carbides particulates such as a W, Mo, Cr, Co, and V in a ferrite matrix, and ~15 wt. % carbides of 2 μm maximum size were measured. This is in good agreement with the microstructural characteristics of similar steels reported in the literature [18,19]. Fig. 1b shows the cross-section of the plasma nitrided M2 steel after chemical etching. The optical image indicates the nitrided layer of ~55 μm which appears as a dark zone after etching with nital reagent (5 vol. %), which corresponds to the compound layer (with thickness ~7.01±0.8 μm) and the diffusion layer, together with a transition zone between the nitrided layer and the steel core. The compound layer is composed of $\gamma'$-$Fe_4N$ and $\epsilon$-$Fe_{2-3}N$ phases in combination with precipitates of nitrides and carbonitrides of alloying elements, which are known to have high hardness and brittleness due to high porosity, nitrogen trapping and decarburization. The diffusion zone is formed mainly of ɣ'-Fe$_4$N phase and M$_6$C carbides and carbonitrides (where M = V, W, Mo, Co) precipitated in the α-Fe ferritic matrix [20].

*Corresponding author: alejandra.garcia@cimav.edu.mx

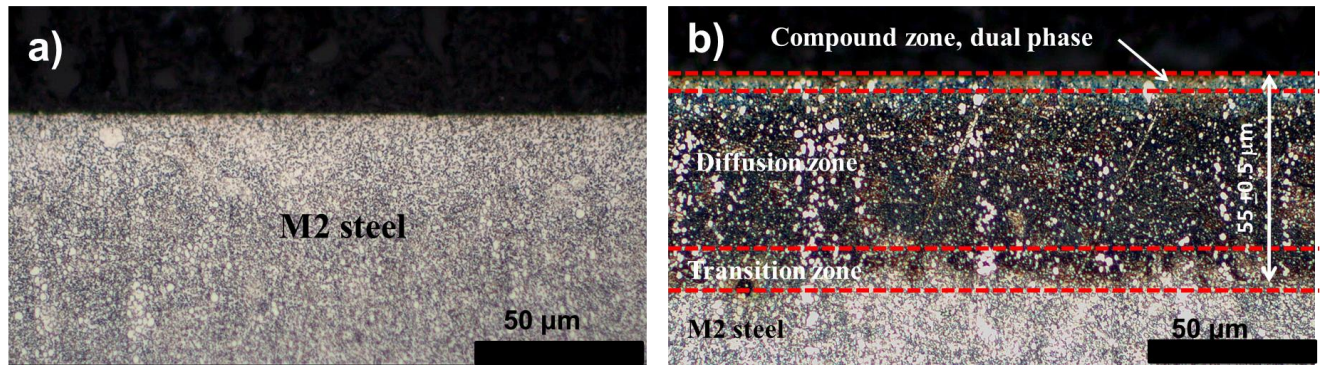

Fig. 1 Optical micrographs of AISI M2 tool steel cross-sections: (a) non-nitrided; b) plasma nitrided.

**3.2 XRD nitrided and non-nitrided substrates.**

**3.2.1 Structural evaluation of nitrided and non-nitrided M2 steel.**

The diffraction pattern of the non-nitrided M2 steel is observed in Fig. 2 (upper pattern), showing a clear evidence of α-Fe ferrite phase with additional diffraction peaks coming from the carbides with hexagonal phases (M$_6$C). This is in agreement with the published literature [21–23], and with the microstructure observations performed in figure 1. Identification of crystalline phases on the PAN sample surfaces are challenging due to nitride formation of the alloying elements and the transformation of primary and secondary carbides into carbo-nitrides or purely nitrides. As the lattice structure of many carbides is similar to that of nitrides, the overlapping of the diffraction peaks occurs, also leading to a increased background level [24]. However, the XRD analysis indicated that iron nitrides ε (Fe$_{2-3}$N) and ɣ'(Fe$_4$N) are present in the compound layer [12,25]. The increase in diffraction peaks broadening and the differences in the position of the peaks indicate the formation of non-uniform stresses within the crystalline structures, primarily due to the inclusion of nitrogen in interstitial sites of the α-Fe lattices and the difference between lattice parameters between the different nitrides and carbo-nitrides formed during the nitriding process [18]. The presence of non-uniform stress is an evidence of second phase segregation hardening, which is responsible for the increase in hardness associated to the plasma nitriding treatment.

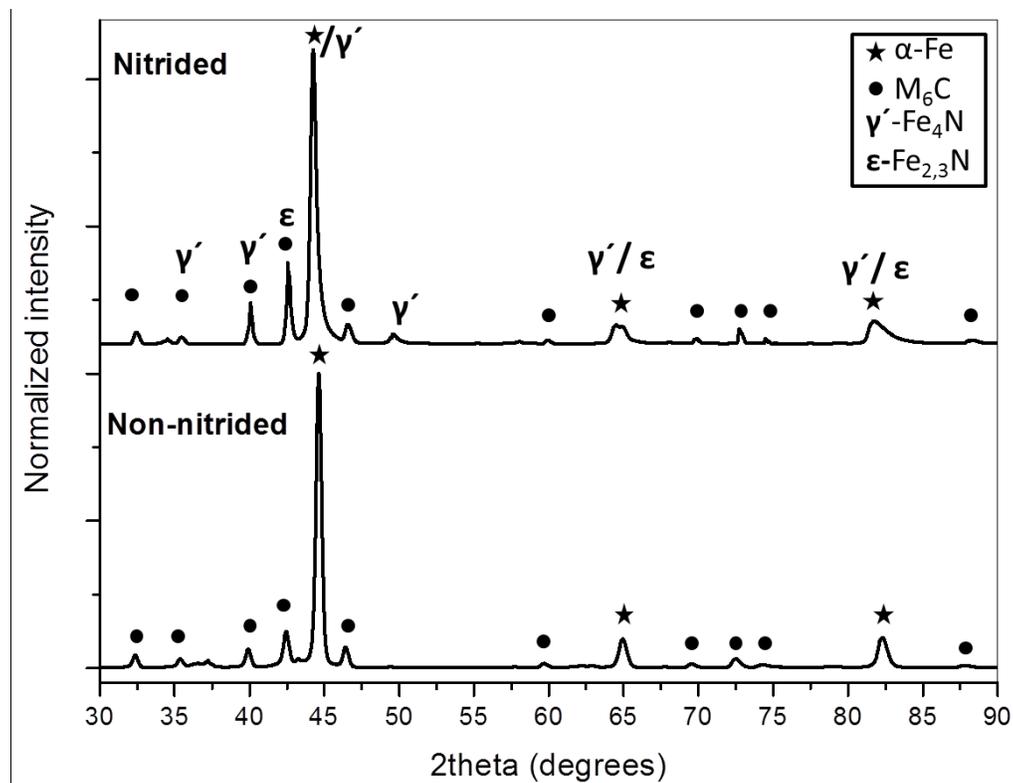

Fig. 2 XRD patterns of nitrided and non-nitrided AISI M2 tool steel.

**3.3 Nitriding case zone**

The nitrided layer was evaluated using Vickers microhardness (HV) measurements under 0.10 N loads (0.01 kgf) on mirror polished cross-sections (Fig. 3). The first indentation was done at 10 μm from the surface with a $HV_{(0.01)}$ of ~762 for nitrided M2 steel which corresponds with the presence of ε-Fe$_2$N and γ'-Fe$_4$N phases along the compound layer. Iron nitrides are considerably harder than the α-Fe matrix [26,27], indicating an increase in hardness above 280% in comparison to the non-nitrided samples, with a HV$_{(0.01)}$ ~ 271 hardness obtained at 30 µm. Hardness profile is in agreement with the nitriding case depth displayed (~ 55 µm) in Fig. 1b. Higher microhardness are shown close to the nitride layer in comparison to the depth diffusion zone, however as discussed in the XRD analysis, this layer is fragile and possess high plastic deformation resistance [28].

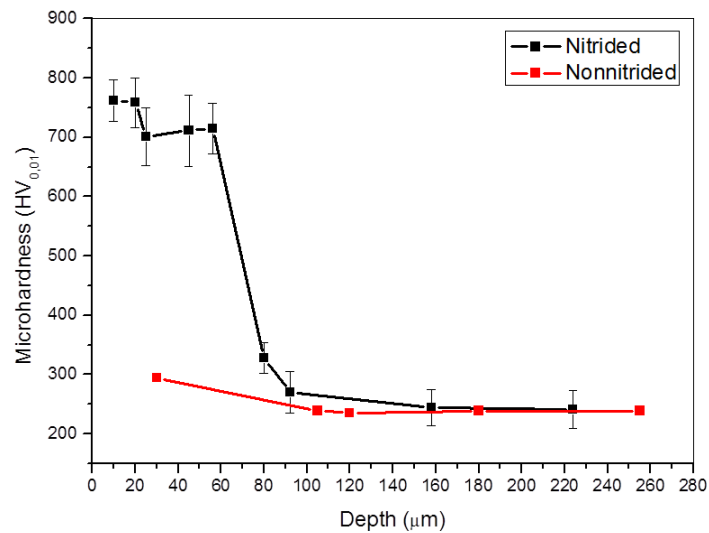

Fig. 3 Microhardness profile along cross-section of nitrided and non-nitrided substrates.

### 3.4 Raman spectra of DLC coatings

The effect of the plasma nitrided surface and PECVD deposition parameters process on the DLC coatings chemistry was studied by Raman spectroscopy (Fig. 4). The synthesized coatings showed the typical DLC Raman spectrum comprising of two carbon signals called D and G-bands. Deconvolution of the Raman spectra using Gaussian functions was conducted to characterize the D and G bands of carbon. The band centered at about 1388 cm$^{-1}$ is associated with $sp^3$ amorphous carbon sites and is called D-band (D standing for "disorder") [29–31]. The characteristic Raman G band was located at 1580 cm$^{-1}$ according with Banerji and Sutton et al. [32,33]. The G peak is associated to the bond stretching of all *sp$^2$* atom pairs in rings structures and line chains [34–36]. In our study, the bias voltages were increased from -380 to -743 V causing the G band shifting from 1588 to 1579 cm$^{-1}$ on the PAN substrate. Similarly, non-nitrided surfaces displayed a G band shift from 1585 to 1571cm$^{-1}$ upon bias voltage increase (Fig. 5a). This Raman shift is an indicative of carbon sp$^2$ type transformation into ring structures. This observation implies an increase of diamond-like bond characteristics of the DLC structure derived from substrate bias voltage increases. In our samples, the DLC-coating with the highest diamond-like component was grown on nitrided substrate. Similar trends were observed by Chang *et.al* [34].

The $I_D/I_G$ ratio measures the disorder degree within the graphitic layer and increases as changes in carbon bonding are induced [33]. Decrease of the $I_D/I_G$ ratio (Fig. 5b), implicate that sp$^3$ bonds transformed into sp$^2$ bonds increasing the graphitization degree [34, 37]. In our results, DLC-coatings on plasma nitrided and non-nitrided surfaces varied from 47 to 61 % according to Fig. 43 inside Robertson report [6], suggesting a DLC a-C:H deposits.

*Corresponding author: alejandra.garcia@cimav.edu.mx

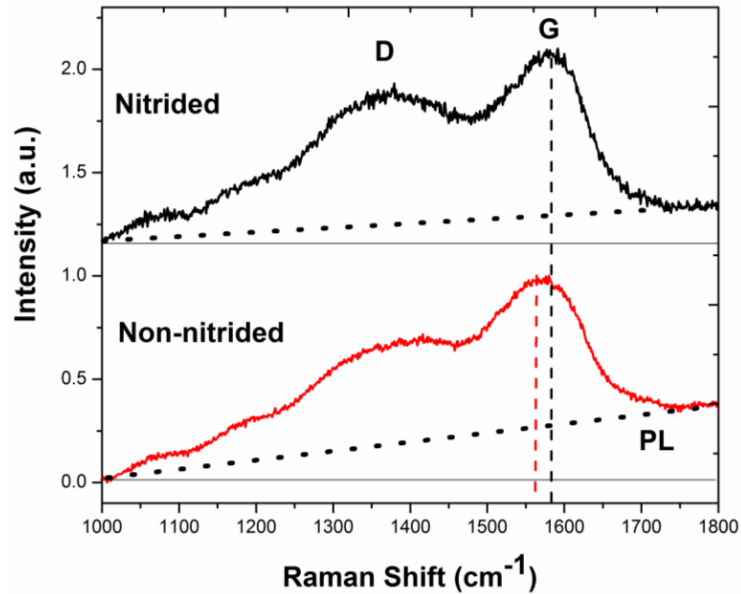

Fig. 4 a) Raman spectrum of DLC (a-C:H) coating deposited on plasma nitrided M2 steel, b) mapping Raman area.

The Raman spectra also provide information on the H content through its photoluminescence (PL) slope [38]. The PL-slope is caused by recombination of electron hole pairs within $sp^2$ bonded clusters in an amorphous $sp^3$ bonded matrix. The PL intensity tends to increase with H concentration increase due to primarily saturation of non-radiative recombination sites (e.g., dangling bonds)[8]. Raman band region between 1000 and 2000 cm$^{-1}$ has shown the PL intensity as a linear scattering background of the PL intensity. By normalizing to the Raman intensity, the PL intensity continues to increase with H content. Normalizing the Raman spectra with respect to the *G* band intensity is reasonable since, neglecting the possible variation of the cross sections along the different DLC-coatings; the Raman intensities still show similar dependence on sample thickness, absorption coefficient, and laser penetration depth. For practical purposes, it is not necessary to measure the PL over a wide spectral range, and we can take the PL background under the D and G band spectra regions [38]. A convenient way to estimate the PL intensity is to ratio the PL background slope to the fitted Gaussian G peak intensity. The slope parameter *S* has a dimension of length, and is best described in micrometer units. The PL intensity caused by hydrogen scattering can be estimated by the slope *S*, which is extremely sensitive to hydrogen content in the film. Equation 1, allows a qualitative evaluation of hydrogen content in DLC coatings [39].

*Corresponding author: alejandra.garcia@cimav.edu.mx

**Equation 1** $\quad H(\%) = 21.7 + 16.6 \log(S/I_G \, [\mu m])$

Hydrogen contents obtained in this investigation (Fig. 5c) was ~ 20 at.% and are in agreement with Fenili *et al*. [40], who obtained a-C:H on carbon steel nitrided and non nitrided. Both DLC deposits on nitrided and non-nitrided substrates showed the same at.% hydrogen which indicate that DLC growth is nondependent of substrate nature. The Raman analysis of results respect to at.% of hydrogen and $sp^3$ content confirm the presence of DLC a-C:H coating for both nitrided and non-nitrated substrates.

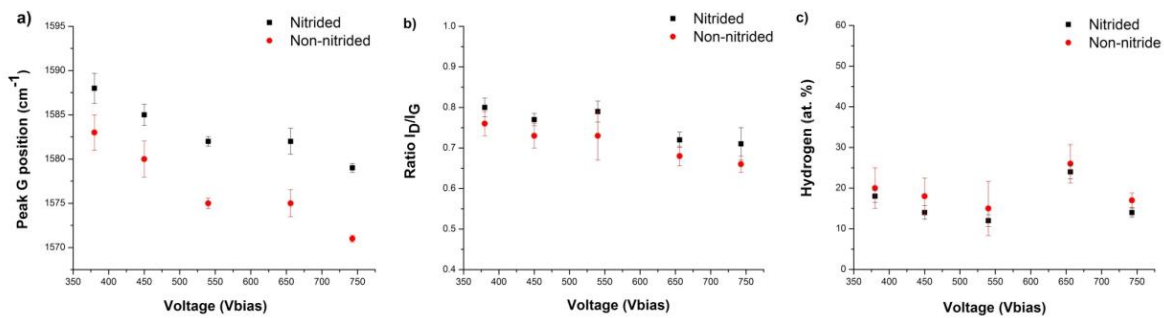

Fig. 5 Raman analysis a) G Peak position, b) $I_D/I_G$ Ratio, and c) Hydrogen content and d) D of the DLC coatings deposited on plasma nitrided and non-nitrided steels, as a function of bias voltage at 150° C.

**3.5. Roughness of the M2 steel and DLC surfaces**

The RMS (root mean square) or $R_q$ roughness of the profile height deviations of the mean line were recorded within the evaluated length. Roughness measurements show that the RMS values increase from 5 nm (*i.e.*, mirror polishing condition) to more than 37 nm due to the PAN process (*e.g.*, ion bombardment and substrate nitrides formation)[1]. A similar tendency is observed for Ra. The DLC-coatings exhibited a uniform surface roughness of about 49 nm on the nitrided substrates, while for non-nitrided samples the roughness was about 8.9 nm. This surface factor may have an impact on the adhesion mechanism of the DLC coatings. The Rq/Ra ratio, known as the coefficient of variation, is useful for describing and comparing different roughness distributions, taking into account that for a symmetric or sinusoidal surface 0 <Rq/Ra <1.11, for a distribution that tends to be Gaussian 1.11 <Rq/Ra <1.25 and for a surface with a random distribution Rq/Ra> 1.25 [41]. As the substrate was mechanically polished, the coefficient of variation had values higher than 1.25 ($R_q/R_a$), with random distributions of roughness and with a slight increase when the DLC-coating is deposited. The coefficient of variation increases when the substrate is nitrided due to the ion bombardment and the formation of nitrides on the surface. This behavior is copied by the DLC film, which has the highest coefficient of variation, with a random distribution of roughness. Roughness distribution are important because during the contact, it will define the initial contact area and will dominate the first stages of the surface behavior in relative movement, in such a way that it will directly influence the resistance to scratching and wear [42].

*Corresponding author: alejandra.garcia@cimav.edu.mx

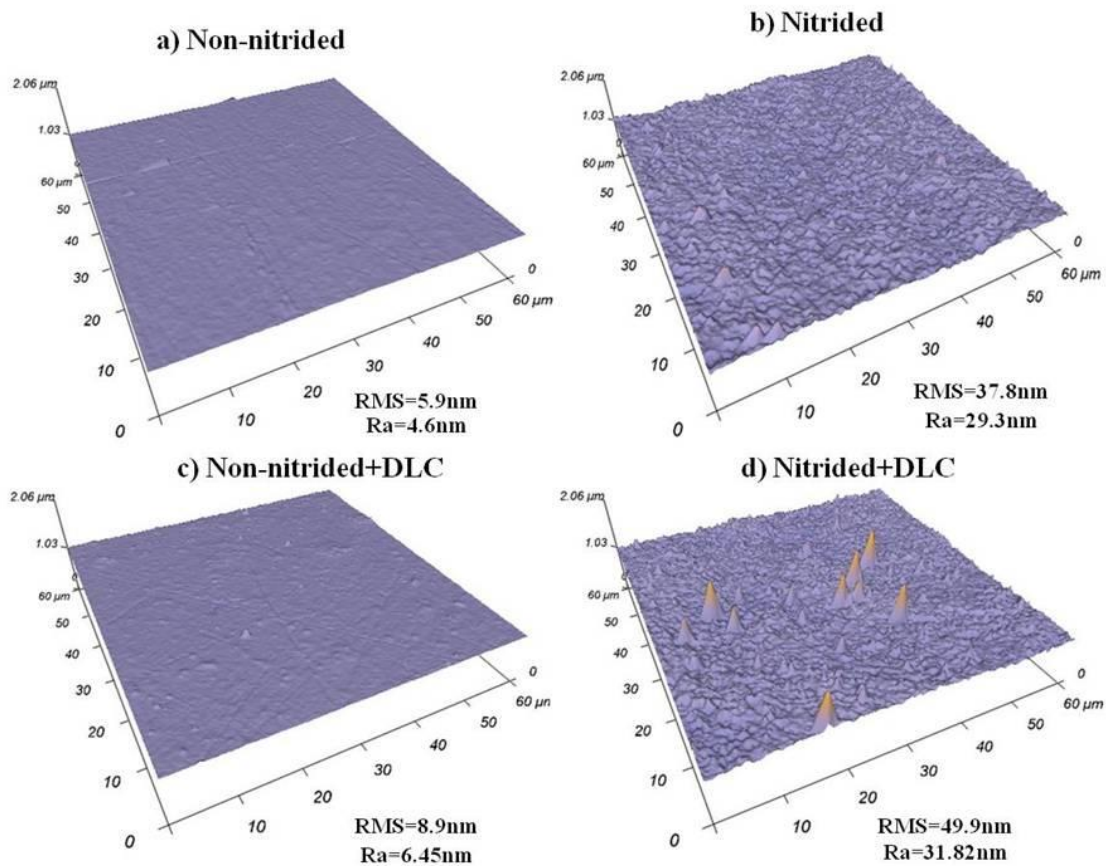

Fig. 6 AFM roughness measurements of a) non-nitrided mirror polished steel, b) plasma nitrided steel, c) non-nitrided surface + DLC, and d) plasma nitride surface + DLC.

**3.6 Adherence behaviors of DLC coatings**

Three scratch tests were done on each coating surface with adjacent tracks separated by minimum 100 μm. Scratch lengths were arbitrarily chosen, while the normal load was steadily raised to ensure coating failure (Fig. 7). Optical microscopy was used to determine the critical loads and evaluate the scratch behavior. Critical loads ($L_{C1}$ and $L_{C2}$) obtained from scratch tests are shown in Fig. 8, according to the ASTM standard C1624-05 [43]. $L_{C1}$ was taken as the load at which cohesive failures (e.g. cracking, chipping of the coating) occurred; $L_{C2}$ was the load corresponding to the onset of adhesive failure (i.e. the load at which the substrate was first exposed). Avelar-Batista *et al.*[44], and Tillmann *et al.*[4] determined a scale including third critical load ($L_{C3}$) from the scratch test, which was defined as the load at which the coating layers were completely removed from the scratch channel. $L_{C3}$ can be taken as a guideline to the maximum load-bearing capacity of different DLC-coated systems, as it represents the critical load-to-failure during scratching at a constant increasing load rate. At this load, a sharp increase in the frictional force result on a fully detached of the coating layers from the scratch track [45]. Scratch tests with increasing normal load were carried out to identify representative scratches for coating adherence as well as to highlight the mechanical response of the DLC. Numerous studies have shown that the system response to a scratch test depends on intrinsic and extrinsic parameters [46] such as substrate, coating, load, loading speed, type of load, diameter and shape of the indenter. These variants have an influence on the failure modes of the coating, and multiple failure modes can be observed [47]. The effect of the contact geometry was studied by Ichimura *et al.* [48] and Xie *et al.*[49] who pointed out that the tip radius highly influences the failure mode of coated solids [45]. Thus indicating that the scratch test conditions are very important if we want compares results with similar works.

Coatings deposited on non-nitrided M2 steel presented a total coating delamination after finishing the scratch test (Fig. 8a). Even the $L_{C1}$ for these substrates was kept in 1N which was the load at the beginning of the test. The critical load $L_{C2}$ is very low and range between 10 and 30 N for non nitrided substrates. The DLC film deposited directly over an M2 steel substrate without any interface layer presents poor adhesion strength values ranging between 20 to 50 N ($L_{C3}$). The as-deposited DLC-coating delaminates right off the substrate surface during testing.

This behavior can be attributed to the high residual tension of the DLC films and the incompatibility in the interface between the DLC reported in the literature [34], in addition, the surface roughness of the mirror polishing of the substrate and the plastic deformation due to the low Steel hardness allow the propagation of interfacial fractures that allow easy delamination of coating (above description corresponds to Fig.9a).

As shown in Fig. 8b and 9b for DLC´s deposited on nitrided samples, at the initial contact stage, the scratch trace is flat and no visible evidence of the film cracking was observed. As the load increases, small length cracks appearing along the scratch trace become visible, but these cracks do not cause appreciable spallation of DLC layer. The DLC-coating begins to fracture when applying the load 10-17 N ($L_{C1}$), although no coating catastrophic failure was detected.

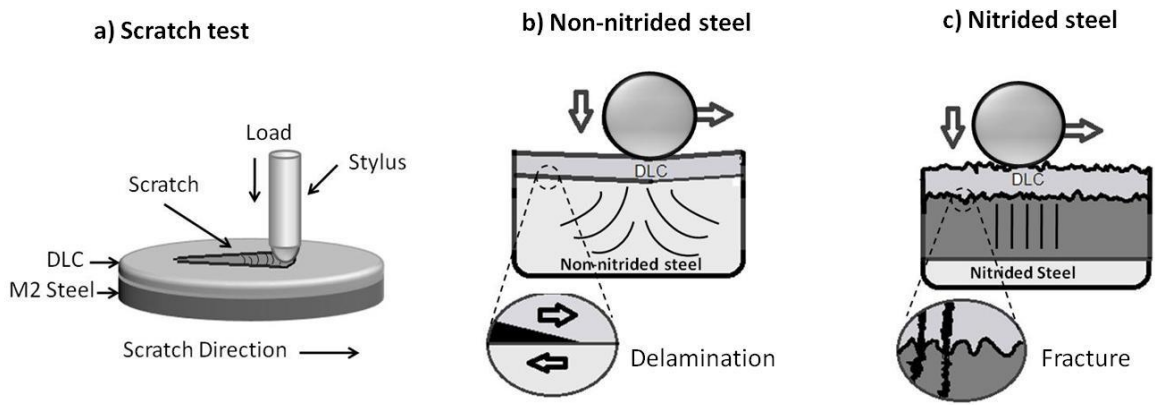

Fig. 7 Failure mechanisms schema in scratch testing of DLC coatings deposited on plasma nitrided and non-nitrided steel.

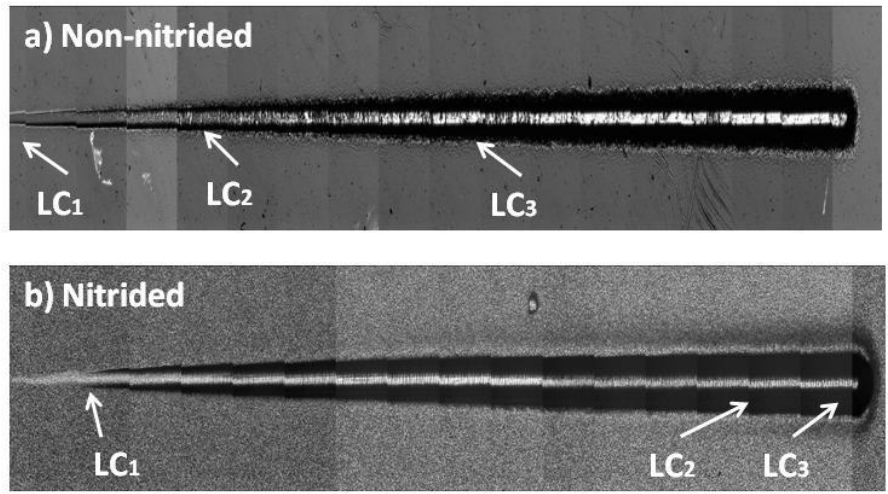

Fig. 8 Critical loads location $L_{C1}$, $L_{C2}$ and $L_{C3}$ in the scratches made on DLCs deposited on plasma nitrided and non-nitrides (100X).

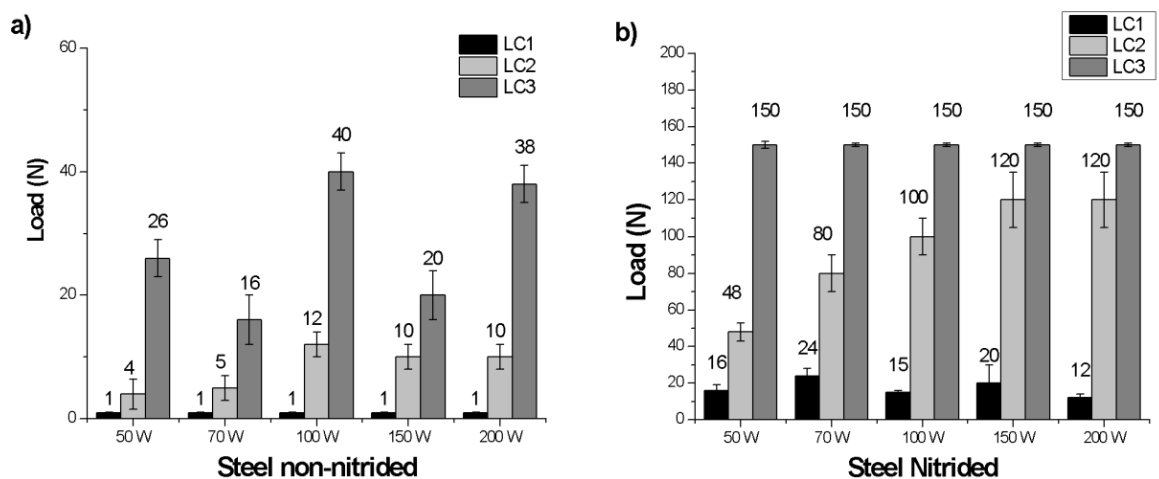

Fig. 9 Critical loads of the DLC coating a) non-nitrided and b) nitrided steel.

On the other hand, $L_{C2}$ values are considerably higher compared to non-nitrided substrates and ranges between 50 and 120 N. The PAN samples coated with DLC by applying 150 and 200 W have the highest $L_{C2}$ values. However, in all samples the coatings withstand after a 150 N, and only partial coating detachment during scratching was detected. For this reason this maximum load value was considered as the $L_{C3}$.

*Corresponding author: alejandra.garcia@cimav.edu.mx

According to our results, a lower $V_{bias}$ gave a higher $sp^2$-fraction. That means that the top layer of the coating system should be more graphitic compared specimens deposited at higher $V_{bias}$, which have favorable effects on the adhesion capability of the coating correlating with results shown in Fig.5b. The DLC-coating on the non-nitrided sample reached only an $L_{C3}$ close to 50 N with the 200 W polarization powers during the deposition.

The nitrided samples show a similar behavior regarding the influence of the $V_{bias}$, despite its surface which is rougher compared to the non-nitrided surfaces. The 50 W deposited sample has a $L_{C2}$ of 48 N and increases to a maximum 120 N at 150 and 200 W. This is consistent with the above mentioned statement that considers the $sp^3$-fraction increase and less graphitization upon $V_{bias}$. The crystalline structure and surface properties of the nitrided layer affect clearly from the amorphous structure of tested coatings, which seems to have a direct effect on the adhesion behavior.

Fig. 10 show the last stage of the scratch by scanning electron microscopy (SEM) image. For coatings on non-nitrided steel (right) the plastic deformation suffered by the steel during the test is observed, the bright region of the worn trace indicates strong gross spallation of the coating. The deformation suffered by the substrate could be one of the causes of the low resistance of the coating adhesion.

In comparison with nitrided M2 substrates, it can be seen that major part of the DLC-coating still remains attached to the surface at the end of the test, with some small failures of the buckling spallation type. However, the increased load brought some enlarged cracks along the scratch trace the catastrophic failure of the films was not reached. This interesting phenomenon may be attributed to several aspects; firstly to the hardness profile of the nitride samples, this avoids the superficial deformation of the substrate and generates greater load support to coating. On the other hand, the roughness prevents the propagation of interfacial cracks and which provides a strong mechanical bond avoiding the interfacial delamination of the DLC. In addition, absorbs the load that produces the scratch test, thus minimizing the coating damage.

*Corresponding author: alejandra.garcia@cimav.edu.mx

*Corresponding author: alejandra.garcia@cimav.edu.mx

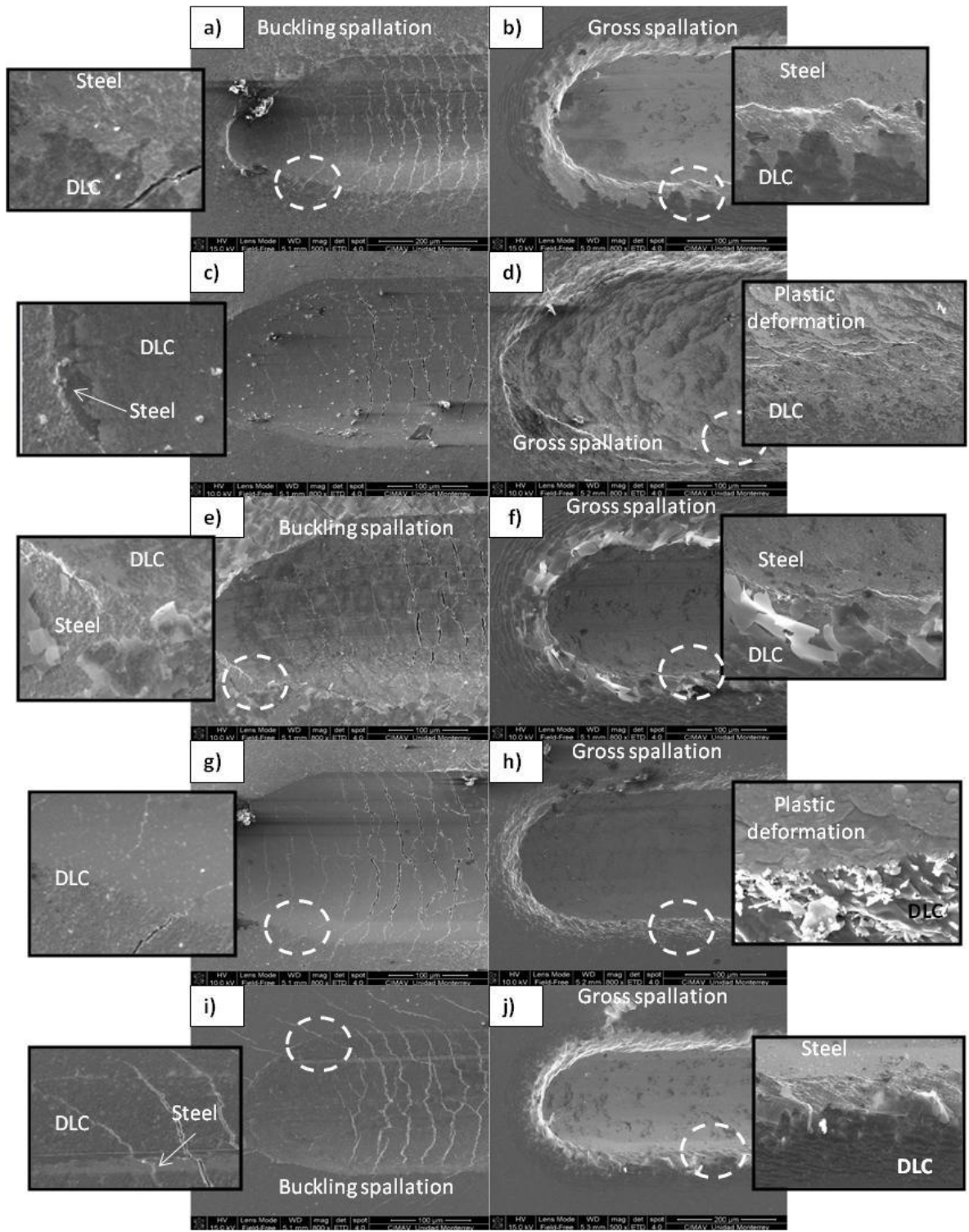

*Corresponding author: alejandra.garcia@cimav.edu.mx

Fig. 10 SEM micrographs of scratch test ends, a) and b) correspond to 50W, c) and d) correspond to 70W, e) and f) correspond to 100W, g) and h) correspond to 150W, i) and j) correspond to 200W of plasma nitrided steels (left) and non-nitrided (right).

## 4. Conclusions

The plasma assisted nitriding effect on the DLC-coating adhesion on M2 steel was investigated. PECVD deposited DLC-coatings were an a-C:H with 20 at% hydrogen and 40-60% sp$^3$ according to Raman analysis. The effect of increasing the power applied during the growth of the DLC coating significantly improved the adhesion on nitrided and non-nitrided M2 substrates. However, $L_{C3}$ values obtained for nitrided samples were three times higher than the non-nitrated ones. Clear differences in the failure mechanism were also observed at the critical load. The DLC coatings on non-nitrided M2 steels suffered catastrophic failures that cause complete delamination of the coatings in a flaking mode. On the contrary, the coatings on plasma nitrided M2 steel mainly showed cohesive failures pictured in internal transverse cracking due to the material hardness.

## 5. Acknowledgments

The financial support by CONACyT though the program "Frontiers of Science" and the project No: 2015-02-1077 is acknowledge. We thank to Luis Gerardo Silva Vidaurri, Miguel A. Esneider Alcala, Nayely Pineda Aguilar, Dr. Francisco Longoria, Dr. Oscar Edgardo Vega Becerra from technical support at CIMAV. Finally, we also thanks the support of the LISMA, CENAPROT and CONMAD infrastructure to conduct this work.

*Corresponding author: alejandra.garcia@cimav.edu.mx

*Corresponding author: alejandra.garcia@cimav.edu.mx

The authors declare that there is not conflict of interest regarding the publication of this paper.